\documentclass{elsart}

\usepackage{epsfig}

\usepackage{bm}       
\usepackage{mathptmx} 

\begin{document}
\begin{frontmatter}

\title { Playing with sandpiles\thanksref{label1}}
\thanks[label1]{
This manuscript has been authored under
contract number DE-AC02-98CH10886 with the U.S.~Department of Energy.
Accordingly, the U.S. Government retains a non-exclusive, royalty-free
license to publish or reproduce the published form of this
contribution, or allow others to do so, for U.S.~Government purposes.
}

\author{Michael Creutz}
\address{Brookhaven National Laboratory,
Upton, NY 11973, USA}

\begin{abstract}
The Bak-Tang-Wiesenfeld sandpile model provdes a simple and elegant
system with which to demonstate self-organized criticality.  This
model has rather remarkable mathematical properties first elucidated
by Dhar.  I demonstrate some of these properties graphically with a
simple computer simulation.
\end{abstract}


\end{frontmatter}

In the mid 1980's Per Bak's office was across the hall from mine.
This was quite fun, with exciting physics always in the air.  One day
Per mentioned that there was a condensed matter seminar that I might
be interested in; so, I went to listen to Chao Tang describe this new
concept of self organized criticality.  I indeed found it quite
fascinating, but not entirely for the right reasons.  At the time I
had been playing with cellular automata on minicomputers, and I
saw that this would give me a new model to play with.

This particular audience is fully familiar with the concept of self
organized criticality, wherein some dissipative systems naturally flow
to a critical state \cite{btw}.  These systems exhibit physics on all
scales, and do this without fine tuning of any parameters, such as the
temperature.

Self-organized criticality provides a nice complement to the concept
of chaos.  In traditional discussions of the latter, one exposes
highly complex behavior arising from systems of only a few degrees of
freedom.  With self-organized criticality one normally starts with
many degrees of freedom, such as the possible locations of grains of
sand in a sandpile, and then extracts simple general features.

The original Bak, Tang, Wiesenfeld paper presents one particularly
simple model to study this phenomenon.  This is a cellular automaton
model formulated on a finite two dimensional square lattice.  On each
site of this lattice is a height variable, a positive integer $Z_i$.
Depending on how you look at it, this variable represents something
between a ``slope'' and a ``height'' for the sand at this point.  If
this variable is too large, i.e. $Z_i>3$, the site is said to be
unstable.  In one time step, all unstable sites undergo a simultaneous
tumbling, reducing the corresponding site by 4 units and adding one to
each nearest neighbor.  With finite and open boundaries, sand spreads
until it is lost at the edges.  Thus repetition of this process will
eventually converge to stability, with all $Z_i < 4$.  One can then
add some more sand and watch the system relax.

As I said above, at the time of these developments I was exploring
simple cellular automata models on the newly appearing inexpensive
microcomputers.  At the time Per was doing similar things; the cover
of the December 1983 issue of Physics Today shows a photograph of Per
Bak's hands in front of a simple Ising simulation on a Commodore 64.
This sandpile model struck me as a natural thing to extend my
programs.  This hobby has continued over the years, and culminated in
a suite of simulation programs for the X window system \cite{xtoys}.
Although they are not as highly developed, this reference also
includes versions for Windows and the Amiga.

\begin{figure}
\centerline{\hbox{
\epsfxsize .33\hsize
\epsffile {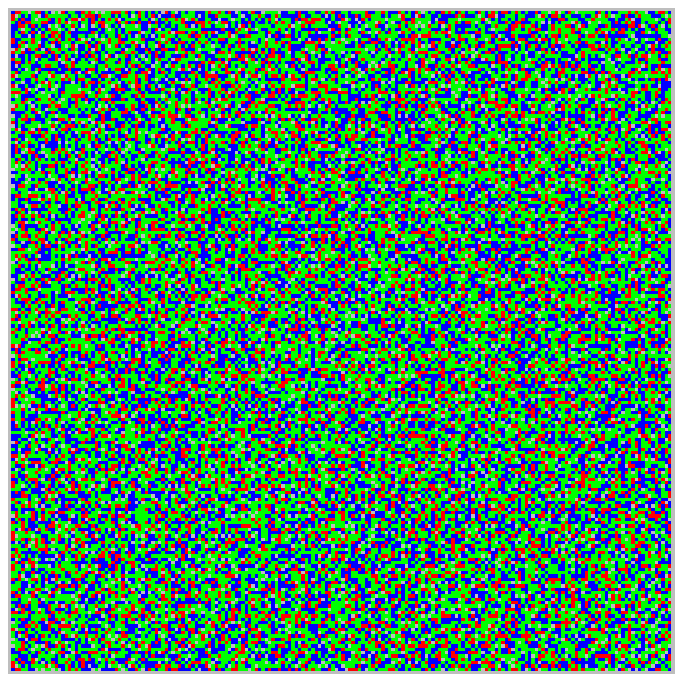}
\hskip .02\hsize
\epsfxsize .33\hsize
\epsffile {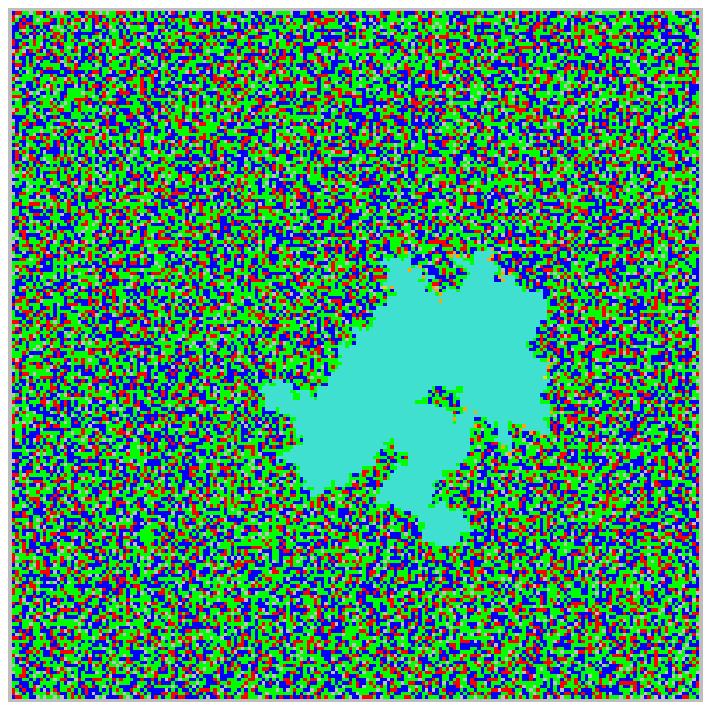}}
\hskip .02\hsize
\epsfxsize .33\hsize
\epsffile {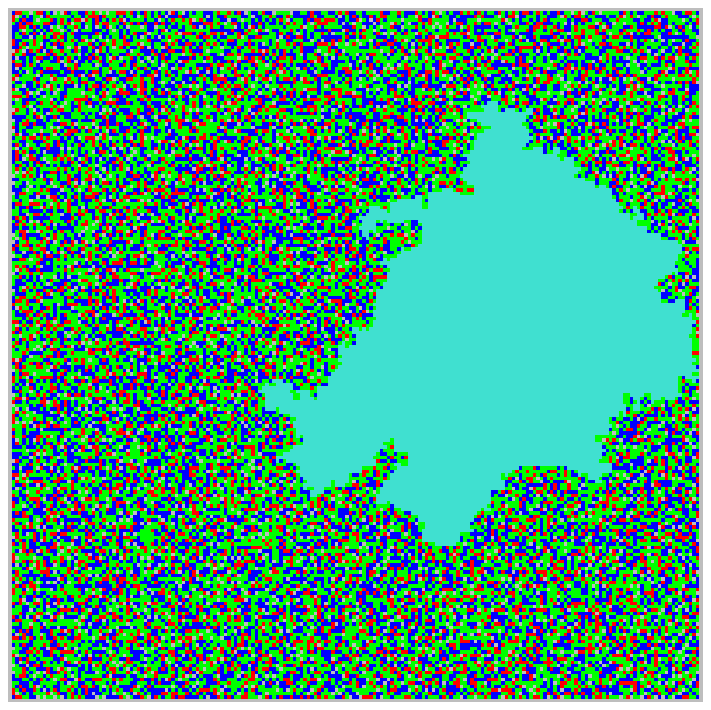}
}
\medskip
\epsfxsize .4\hsize
\centerline {\epsffile{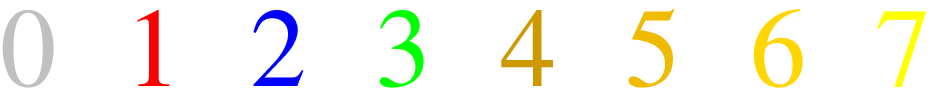}}
\caption{The progress of an avalanche on a typical critical sandpile
configuration on a 198 by 198 lattice.  The light blue region is where
the avalanched has progressed.  The first image is the initial state,
the second while the avalanche is underway, and the final shows all
sites that have tumbled during the full relaxation process.  The color
code for the site heights
appears
below the figure.
}
\label{avalanche}
\end{figure}

With these programs one can do the classic avalanche experiment live
on a computer screen.  Figure \ref{avalanche} shows a critical initial
state, an active avalanche, and the region covered by the avalanche
after it stops.

After I had played with this model for a couple of years, Deepak Dhar
produced some rather remarkable results on the analytic properties of
the model \cite{dhar}.  The above programs allow a rather elegant demonstration of
some of these results.  

To make things precise, let me begin with some definitions.  A
``configuration'' $C$ for the sandpile model is a set of integer
heights $Z_i$ on the sites $i$ of a two dimensional finite lattice.
Associated with each site is a ``tumbling operator'' $t_i$.  When
applied to a configuration $C$ this operator reduces $Z_i$ by four
units and adds one to each of its neighbors.  This is characterized by
a ``tumbling matrix'' $\Delta_{ij}$ which indicates how much site $j$
changes with a tumbling at site $i$.  Thus $t_iC$ is a new
configuration with modified heights
$$
Z_j\rightarrow Z_j-\Delta_{ij}
$$
For the simple nearest neighbor model
$$
\Delta_{ij}\ =\
\left\{
\matrix
{
& 4 & \hfill i=j \cr
& -1 &\qquad\hfill i,j\ {\rm neighbors}\cr
& 0 &\hfill {\rm otherwise}\cr
}
\right .\qquad
$$

We then define the relaxation process by tumbling all sites with
$Z_i>3$ and then repeating this to stability.  Stability will always
be achieved eventually since in the process sand spreads towards the
boundaries, making the total amount of sand monotonically decrease.
Applying sand to a stable system is formalized with the definition of an 
``addition operator'' $a_i$ so that $a_iC$ is a new configuration
obtained by taking $Z_i\rightarrow Z_i+1$ and then relaxing.

Note that after dumping lots of sand to the system, some stable
configurations cannot be reached.  For example, one can never make two
adjacent two adjacent $Z_i=0$.  This is because in trying to tumble
one to zero, the neighbor gains a grain, and vice versa.  This leads
to the definition of the ``recursive set'' $R$, which consists of any
stable configuration that can be obtained by adding sand to any state.
This set is not empty since one can always reach the ``minimally
stable state'' $C^*$, defined by having all $Z_i=3$.

With these definitions at hand, I now introduce two remarkable
theorems proved by Dhar \cite{dhar}.  The first is that the the $a_i$
commute
$$
a_ia_jC=a_ja_iC
$$ 
The proof uses the linearity of the $t_i$.  In the relaxation
processes represented by the two sides of this equation, the order of
tumblings can be rearranged, but the final configurations are equal.
This Abelian nature is reminiscent of the process of long addition.
In this analogy the tumbling process is like carrying.

The second theorem is that if we restrict ourselves to the recursive
set, then the operator $a_i$ is invertible.  More precisely, suppose I
am given given some configuration $C$ which is in the recursive set
$R$.  Then for any specified site $i$ there exists a \underbar{unique}
configuration $C^\prime$ also in $R$ such that $a_iC^\prime=C$.  Thus
we say that $C^\prime\equiv a_i^{-1}C$.

Dar showed that these theorems have some interesting immediate
consequences.  One is that we can now characterize the ``critical
ensemble'' as being an ensemble of recursive states where any given
recursive state is equally likely as any other.  Also, the number of
recursive states is simply the determinant of the toppling matrix
$\vert\Delta\vert$.  For a large system of $N$ sites this determinant
grows as $\vert\Delta\vert\sim (3.2102\ldots)^N$. Thus the recursive
states become a set of measure zero in comparison to the $4^N$ total
stable states.

\begin{figure}
\epsfxsize .35\hsize
\centerline{\epsffile{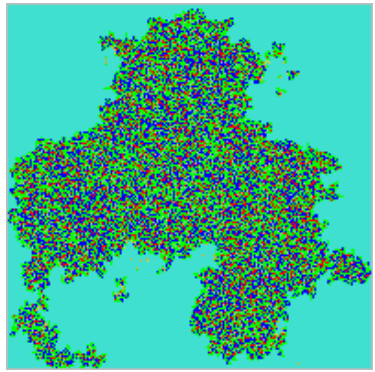}}
\caption{A critical configuration in the process of undergoing the
burning algorithm.  Dropping one grain of sand from each open edge
causes an avalanche that tumbles every site exactly once.  A state not
in the critical ensemble will leave some sites unburnt.}
\label{burn}
\end{figure}

Dhar later provided a simple algorithm to determine if a configuration
is recursive.  This ``burning algorithm'' starts with adding one sand
grain to $C$ from every open edge.  This is easily implemented by
turning on ``sandy boundaries'' for one time step.  On a rectangular
system, the process dumps one grain on each edge site and two on
corner sites.  After this addition, the system is allowed to update to
stability.  Then the original configuration $C$ is recursive if and
only if each site tumbles exactly once during the relaxation process.
Figure \ref{burn} shows this process in action for a representative
recursive state.  In this figure light blue indicates those sites that
have already tumbled.  On completion, the entire lattice acquires this
color.

The burning algorithm leads to an amusing result on sub-lattices.  Consider
extracting from a lattice an arbitrary connected sub-lattice.  If on
this sub-lattice we set the sand heights to their corresponding values in some
recursive state on the full lattice, then the resulting configuration
is recursive on the subset.  This follows since in the burning of the
large lattice, topplings on sites just outside the sub-lattice will
dump exactly the amount of sand on the sub-lattice as required to start
the burning algorithm there.  Then this will all burn, just as if it
was on the larger lattice.

But this result has the deeper consequence that any avalanche started
on a recursive state by any addition of sand will be simply connected.
If an avalanche region is not simply connected, i.e. it has an island
of untoppled sites, then the topplings outside this region would have
created just the dumping on the region necessary to burn it.  This
explains the absence of any untumbled islands in the final avalanche
region of Fig. \ref {avalanche}. It is an amusing game to take a
recursive state and add sand in attempts to create an untumbled region
bounded on all sides by a tumbled region.  Somehow the system knows
that this is not allowed.  But take away some sand so as to exit the
recursive set, and then creating untumbled islands becomes easy.

There is a natural mapping between the group generated by the addition
operators $a_i$ and the recursive set itself.  Indeed, we easily can
define an operation of addition between configurations.
Given two stable configurations $C$ and $C^\prime$
with slopes $Z_i$, $Z_i^\prime$ respectively, we 
define $C\oplus C^\prime$ by relaxing $Z_i+Z_i^\prime$.
Under $\oplus$ the recursive states form an Abelian group

This raises another amusing question \cite{mysand}.  Since we have a
group, we must have an identity state.  What is the configuration
representing this state $I$?  This state must both be recursive itself
and have the property that $I\oplus C=C$ if and only if $C$ is in the
recursive set.  Indeed, it is the unique nontrivial configuration with
$I\oplus I=I$.  It cannot be the empty state, since that is not
recursive.

\begin{figure}
\centerline{\hbox{
\epsfxsize .33\hsize \epsffile{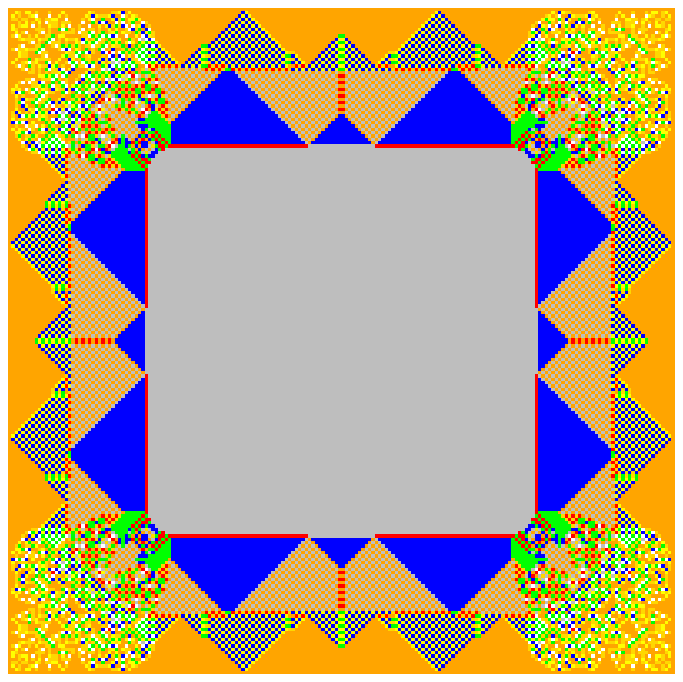}
\hskip .02\hsize
\epsfxsize .33\hsize \epsffile{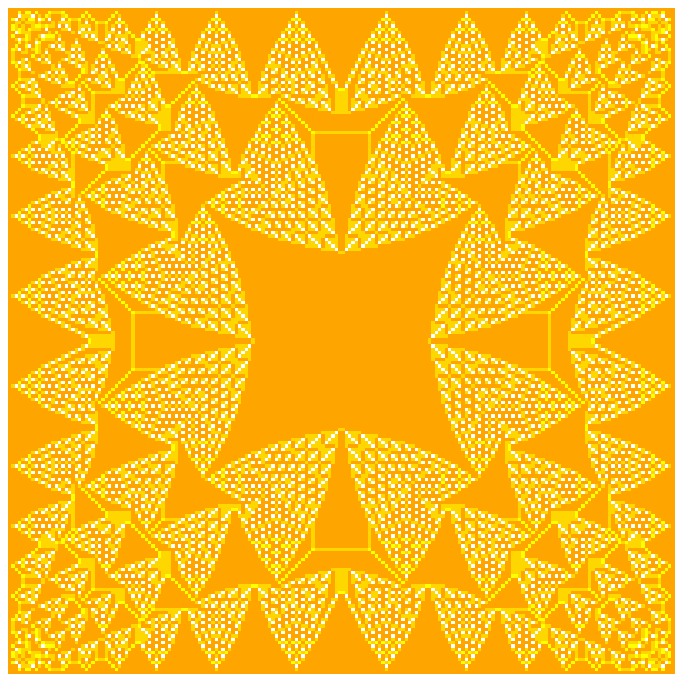}
\hskip .02\hsize
\epsfxsize .33\hsize \epsffile{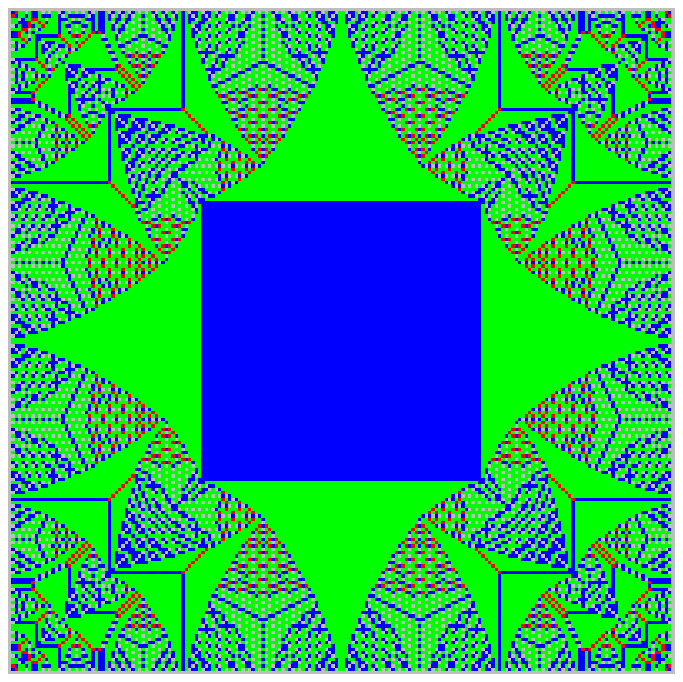}
}}
\caption {Constructing the identity state.  An initially empty state is
run with sandy boundaries giving an inflow as in the first image.
After the system with the sandy boundaries stops changing, we have the
situation in the middle.  Then, on switching to open boundaries, sand
runs off to leave the identity state shown in the third image.}
\label{identity}
\end{figure}

A simple algorithm to
construct $I$ follows from the identity
$$
a_i^4=\prod_{\rm neighbors}a_j
$$
This follows since adding four grains of sand to any site will force a
tumbling.
Combining this with the fact that
$a_i$ invertible on recursive set gives the identity
$$
1=a_i^4\prod_{\rm neighbors}a_j^{-1}
$$
If we multiply this equation over all sites, the $a_i$ factors cancel
on interior sites.  But one power of $a_i$ is left on each edge and
two grains of sand are added on each corner.  Indeed, this product
forms the basis of the burning algorithm.  From this construction we
find a configuration, call it $I_0$, with one grain of sand on each
edge and two on each corner.  This configuration has the property that
if added to a recursive state it leaves that state unchanged, i.e. if
and only if $C$ is recursive
$$
I_0\oplus C=C
$$
However $I_0$ is not itself recursive since it has lots of empty sites
in the middle.  Therefore $I_0$ is not the desired $I$.  To find the
latter, we can simply iterate the above process.  For this start with
and empty table, make the boundary conditions sandy, and run until the
table fills up.  Then go back to open boundaries and relax back to
stability.  The final state is then the desired identity
configuration.  This process is sketched in Figure \ref{identity}.

Hopefully I have convinced you that this simple sandpile model is lots
of fun to play with.  The results I have mentioned are just a few of
many.  Some simple additional properties include the fact that if $C$
recursive, then in constructing $C\oplus I$ the number of topplings at
any given site is independent of $C$.  Also, a single added grain $n$
sites from the edge can tumble no site more than $n$ times.  Finally a
single grain added anywhere can tumble a site $n$ steps from the edge
no more than $n$ times.

\vfill\eject

\end{document}